\documentclass[aip,rsi,reprint,graphicx]{revtex4-1} 

\usepackage{graphicx}%
\usepackage{dcolumn}%
\usepackage{bm}%
\usepackage{epstopdf}
\usepackage{amssymb}

\begin{document}

\title{Cryogenic coaxial microwave filters}

\author{G. Tancredi}
\thanks{G.Tancredi and S. Schmidlin contributed equally to this work}
\affiliation{Department of Physics, Royal Holloway, University of London, Egham, TW20 0EX, UK}
\author{S. Schmidlin}
\thanks{G.Tancredi and S. Schmidlin contributed equally to this work}
\affiliation{Department of Physics, Royal Holloway, University of London, Egham, TW20 0EX, UK}
\affiliation{National Physical Laboratory, Hampton Road, Teddington, TW11 0LW, UK}
\author{P. J. Meeson}
\affiliation{Department of Physics, Royal Holloway, University of London, Egham, TW20 0EX, UK}

\begin{abstract}
The careful filtering of microwave electromagnetic radiation is critical for controlling the electromagnetic environment for experiments in solid-state quantum information processing and quantum metrology at millikelvin temperatures. We describe the design and fabrication of a coaxial filter assembly and demonstrate that its performance is in excellent agreement with theoretical modelling. We further perform an indicative test of the operation of the filters by making current-voltage measurements of small, underdamped Josephson junctions at 15~mK.
\end{abstract}

\maketitle

Recent breakthrough results in the fields of quantum information processing and quantum metrology have largely been attributed to the extreme care and thought put into experimental design. In particular, it has been critical to isolate the devices under test from the electromagnetic environment. Whilst these experiments require low temperatures of $\lesssim50\,$mK, the electron temperatures of the devices are often found to be greater than the bath temperature, as a result of uncontrolled electromagnetic noise entering the system and propagating along the cryostat wiring. It is often assumed that the primary noise source of the measurement setup at 300~K is Johnson noise, which has a white spectrum extending up to $\frac{k_{B}T}{h}\simeq\,6\,$THz. Thus an effective cryogenic filter should reject noise between $6$~THz and $1$~GHz, the latter corresponding to a temperature of $50\,$mK, and it should be placed at the lowest temperature stage in order to decrease the bandwidth and intensity of its own Johnson noise.
 
Since commercial filters are not adequate to screen up to the THz range at millikelvin temperatures, several experimental groups have designed and implemented different types of microwave cryogenic filters\cite{Martinis1987,Milliken2007,Lukashenko2008,Bluhm2008,Vion1995,leSuer2006,Zorin1995}. These filters are probably comparable in performance and the preferences between them are often determined by factors such as the space available within the experimental set-up or the ease of manufacture. 

We designed, built and tested robust filters, based on commercially available semi-rigid stainless steel coaxial cables. We show that such filters have a highly predictable performance. A section of dissipative coaxial cable provides a simple and effective means with which to fabricate a filter with predictable attenuation and bandwidth, as both of these quantities depend only on the physical and geometrical properties of the cable\cite{Zorin1995}.
 \begin{figure}
\begin{centering}
\includegraphics[trim=0cm 0cm 0cm 0cm,clip=true,scale=0.97]{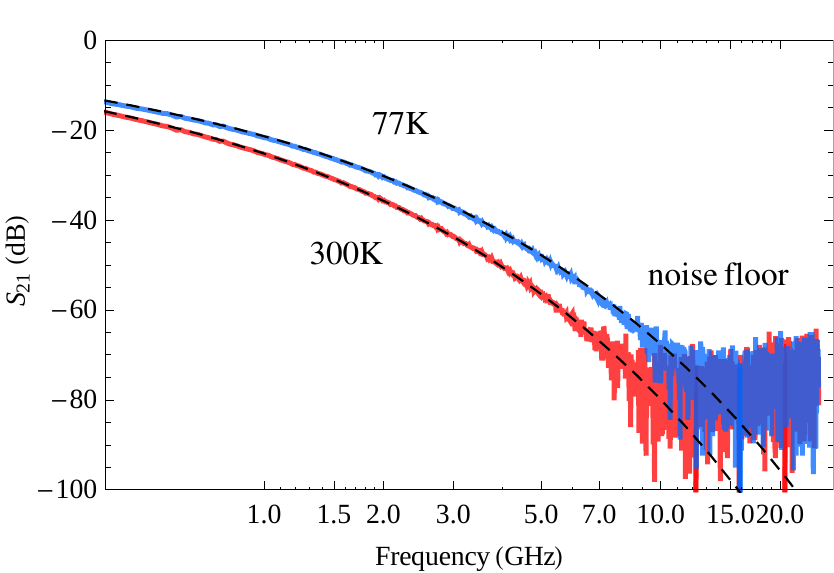}
\caption{\label{fig:transmission} The forward scattering parameter $S_{21}$ as function of frequency for a single 1.5~m long coaxial cable, wrapped around the copper cylinder, with SMA connectors mounted at each ends. The continuous and dashed lines represent the measurements of a VNA and the fit given by Eq.\ref{A} with $\rho$ as fitting parameter.} 
\par\end{centering}
\end{figure}

Our filters consist of a length of semi-rigid coaxial cable (UT-020-SS-SS\cite{Microcoax}). This cable has a 304 stainless steel jacket (outer diameter $d=0.508$~mm, inner $d_{2}=0.381$~mm) and inner core ($d_{1}=0.114$~mm). The mediating dielectric is PTFE. For proper functioning of the filters the entire length of these cables needs to be well thermalised to the mixing chamber temperature. In our design four 1.5 m long cables are wrapped around and soldered to a hollow copper cylinder, open on one end, of length 30~mm and outer diameter 40~mm, with a wall thickness of 2~mm. This arrangement allows simultaneous filtering of four measurement wires.
The attenuation per unit length $\mathcal{A}(\omega)$, provided by the coaxial cable, at a frequency $\omega$ can be calculated following the theoretical model by Zorin\cite{Zorin1995}. $\mathcal{A}(\omega)$ is related to the propagation coefficient $\gamma(\omega)=\sqrt{i\omega c (r+i\omega l)}$ by:
\begin{equation}\label{A}
\mathcal{A}(\omega)=20 \Re\gamma(\omega)/\ln(10).
\end{equation} 
$c$, $l$ and $r$ represent the coaxial cable capacitance, inductance and resistance per unit length respectively:
\begin{equation}\label{c}
c=2\pi\epsilon\epsilon_{0}/\ln(d_{2}/d_{1}),
\end{equation} 
\begin{equation}\label{l}
l=\mu\mu_{0}\ln(d_{2}/d_{1})/(2\pi),
\end{equation} 
\begin{equation}\label{r}
r=\sqrt{\frac{\omega\mu\mu_{0}}{2}}\,\left(\frac{\sqrt{\rho_{c}}}{\pi d_{1}}+\frac{\sqrt{\rho_{j}}}{\pi d_{2}}\right).
\end{equation} 
$\epsilon$ and $\mu$ are the dielectric constant and the magnetic permeability of the dielectric. $\rho_{j}$ and $\rho_{c}$ are the resistivity of the jacket and core material (in our case $\rho=\rho_{j}=\rho_{c}$, $\epsilon=2.1$ and $\mu=1$). Given the geometry and construction materials of our cables, we expect  $\mathcal{A}\left(1\,GHz\right)\simeq-17$~dB/m at room temperature with $\rho_{300K}=7.19\times10^{-7}\,\Omega\,$m. For comparison, the estimated value of $\mathcal{A}\left(1\,GHz\right)$ for a $0.5$~mm Thermocoax cable is $\simeq-42$~dB/m. Thus for our coaxial cable of length $l_{0}=1.5$~m we expect an attenuation $A\left(1\,GHz\right)=l\times\mathcal{A}\left(1~GHz\right)\simeq-25$~dB. It is important to stress that this model ceases to be valid at high frequency when the wavelength becomes comparable to the cross section of the cable. This means the above equations should be a good predictive model up to $\simeq1$~THz, assuming the material dependent parameters remain predictable.

Before installing the filters in our dilution refrigerator, we measured the forward scattering parameter $S_{21}\left(\omega\right)$ as a function of frequency at 300~K and 77~K with a Vector Network Analyzer (VNA). Several thermal cycles to 77~K were performed to check the mechanical strength of the filter and to ensure repeatable behaviour. Fig. \ref{fig:transmission} shows $S_{21}$ measured at 300~K and 77~K for a single coaxial cable, wrapped around the cylinder, with an SMA connector mounted at each end. Using Eq.\ref{A}, we fit the experimental data, up to 7~GHz, with $A\left(\omega\right)=l_{0}\times\mathcal{A}(\omega)$ using the resistivity $\rho$ as the only fitting parameter. We obtain $\rho_{300K}=\left(7.22\pm0.12\right)\times10^{-7}\Omega\,$m at 300K and $\rho_{77K}=\left(5.27\pm0.04\right)\times10^{-7}\Omega\,$m at 77~K. At 300~K, the electrical resistivity $\rho_{300K}$ is compatible with previous measurements while, at 77~K, $\rho_{77K}$ is $\simeq3\%$ greater than the expected value of $5.1\times10^{-7}\,\Omega\,$m\cite{White2002}. These fits highlight one advantage of the design, their performance is highly predictable at the design stage. The resistivity of stainless steel remains essentially unchanging below 77~K, thus the behaviour of the filters when in use below 4 K may be predicted with confidence.

\begin{figure}
\begin{centering}
\includegraphics[trim=1.5cm 11cm 20cm 0cm,clip=true,scale=0.62]{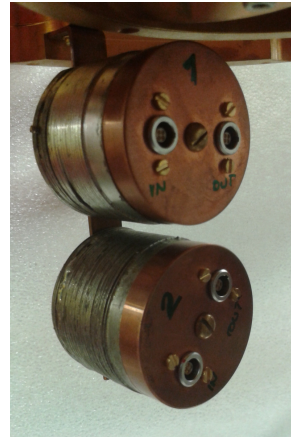}
\caption{\label{fig:filter} A view of the filters as finally assembled on the mixing chamber of our refrigerator. The Microcoax is soldered to the outer of the copper cylinder while the ends enter through two small holes into the inner cavity where they connect to the Lemo connectors.}
\par\end{centering}
\end{figure}

Figure~\ref{fig:filter} shows two filters installed at the mixing chamber stage of our dilution refrigerator. The filter construction process is as follows. Due to the presence of an oxide layer on the stainless steel, the coaxial cables were carefully prepared for installation following well known soldering techniques. The jacket of each cable was abrasively cleaned and degreased with acetone and its surface was wetted with Superior No 23 flux\cite{flux}, specially formulated for soldering stainless steel. The surface was then progressively tinned along its length with multicore 60/40 tin-lead solder using a soldering iron with a temperature of $200^{\circ}$~C. The soldering temperature must not be greater as we discovered the PTFE may expand and the outer stainless steel jacket may burst. The four tinned cables were then wrapped around the copper cylinder, which had previously also been solder coated. The coaxial cables were mechanically fixed to the cylinder using metallic clamps (with PTFE protective pads to avoid squeezing and damaging of the outer jacket). The cylinder and tightly wound cables were then placed on a temperature-controlled hot plate at a temperature of $200^{\circ}$~C allowing the pre-coated solder to melt. Additional flux and solder were carefully applied. After careful cleaning the eight stainless steel cable ends were passed through two holes into the cavity of the copper cylinder and the cores were soldered into a 4-pin panel-mounted Lemo connector, ad-hoc brass cable supports kept the coaxial cables in place to avoid stressing and breaking the fine inner wire. The cable supports and the Lemo connectors were fixed into a copper plate that forms an end-cap for the copper cylinder. The cable cores were soldered and the cables glued to the supports using double-bubble epoxy. Once dried, the cover was fixed to the main cylinder via a copper rod which is simultaneously used to thermalise the filter assembly to the mixing chamber. This arrangement allows complete screening of the inner wire of the cables from electromagnetic radiation. In this assembly we wished to use Lemo rather than SMA connectors for connecting to existing sample boxes, however, this created short ($\approx 1$ mm) sections of unshielded inner wire. Consequently there is the possibility of short-circuiting the filter by RF modes of frequency 1- 10 GHz in the copper cavity, though the estimated coupling is small and no evidence for such short-circuits were seen in testing. 

\begin{figure}
\begin{centering}
\includegraphics[trim=5cm 17.5cm 0cm 0.5cm,clip=true,scale=0.8]{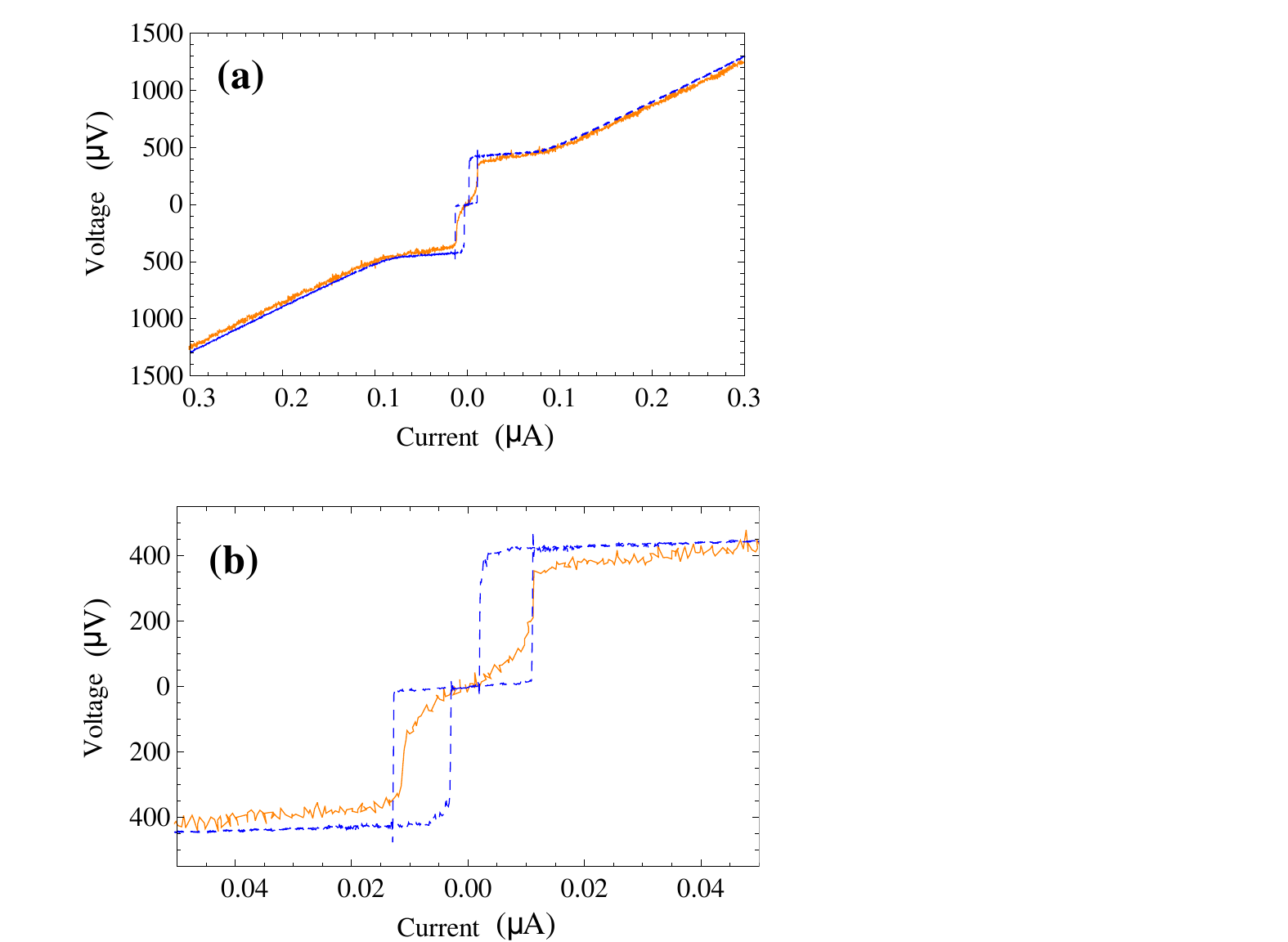}
\caption{\label{fig:IV} \textbf{(a)}. Voltage-Current characteristic of two Josephson junctions at 15~mK. The dashed(solid) line represents the V(I) curve of sample JJ2(JJ1), measured with(without) the coaxial cable filter installed at the mixing chamber stage. \textbf{(b)}. Expanded view of \textbf{(a)}.}
\par\end{centering}
\end{figure}
In order to further verify the operation of our filters we measured the voltage-current characteristic of sub-micron sized Al/AlOx/Al junctions at $\simeq15$~mK with and without the filters inserted. The junctions were  fabricated on a Si/SiO$_{2}$ substrate with overlap areas of $\simeq250\times250\,$nm$^{2}$ and a normal state resistance of $R_{N1}=(3.8\pm0.05)$k$\Omega$ (sample JJ1) and $R_{N2}=(3.9\pm0.05)$k$\Omega$ (sample JJ2).

A Josephson junction in a quiet environment and in the limit of small damping displays an hysteretic behaviour in its voltage-current or V(I) characteristic. Interactions with high energy photons (with frequency above $\Delta/h\simeq50$~GHz, with $\Delta$ the superconducting energy gap), from the leads or incident from the vacuum, may drastically affect the ``ideal" V(I) curve. For example they may cause the break up of Cooper pairs which can switch the junction state from the superconducting to the resistive branch at bias currents below the critical current $I_{c}$ as well as causing phase diffusion and other effects. Hence by inserting the filters at the lowest temperature stage we attempt to limit the Johnson noise arising from the 300~K experimental setup that extends up to about $6$~THz.

We performed two-terminal V(I) measurements by current biasing the junction and measuring the voltage across it after amplifying the signal with a low-noise amplifier. Fig.~\ref{fig:IV}\textbf{(a)} shows the V(I) characteristic of sample JJ1 (continuous line) and sample JJ2 (dashed line). Sample JJ1 was measured with no filters in place while JJ2 was measured with the filters installed at the mixing chamber stage. Fig.~\ref{fig:IV}\textbf{(b)}, shows an expanded view for smaller current values and provides a better understanding of the action of the filters. The incorporation of the filters completely changes the shape of the V(I)-curve below the critical current. In addition, we observe that the voltage noise in the resistive branch was reduced. Without filters the sample does not display a supercurrent branch at 0~V while this is now clearly visible for JJ2 with filters inserted. JJ2 also displays a sharp transition from the supercurrent branch (V= 0~V) to the resistive state branch, with the junction developing a voltage of $2\,\Delta/e\simeq0.365$~mV at low current values, which agrees with theoretical prediction. For JJ1 this transition is not sharp and the hysteretic behaviour typical of an underdamped junction is not visible. 

We have designed, built and tested robust, predictable and simple to construct low-pass cryogenic filters intended for quantum information processing and quantum metrology experiments. The attenuation achieved is in excellent quantitative agreement with the theoretical model proposed by Zorin\cite{Zorin1995}.  We have shown that the current-voltage characteristic of Al/AlOx/Al nano-scaled Josephson junctions improves with the addition of these filters to the measurement circuit, confirming that the electrical environment plays a crucial role even in such simple experiments.
\begin{acknowledgments}
The authors wish to thank G. Ithier, N. Langford, J. Pekola and I. Wisby for useful discussions and R. Elsom and M. Venti for the technical support. This work was supported by the European Community FP7 Programme under Grant Agreement No.228464 (MicroKelvin, Capacities Specific Programme) and by the EPSRC under Grant Number EP/F041128/1.
\end{acknowledgments}

\end{document}